\documentclass[aps,prb,reprint,showkeys,superscriptaddress,floatfix]{revtex4-1}
\usepackage{amsfonts}
\usepackage{amssymb}
\usepackage{amsmath}
\usepackage{hyperref}
\usepackage{graphicx}
\usepackage{bookmark}
\begin{document}
\title{Wetting Properties of Graphene Aerogels}
\author{Francesco De Nicola}
\email[E-mail: ]{francesco.denicola@iit.it}
\affiliation{Graphene Labs, Istituto Italiano di Tecnologia, Via Morego 30, 16163 Genova, Italy}
\author{Ilenia Viola}
\affiliation{CNR NANOTEC-Institute of Nanotechnology, S.Li.M. Lab, Department of Physics, University of Rome La Sapienza, P.le A. Moro 5, 00185 Roma, Italy}
\author{Lorenzo Donato Tenuzzo}
\affiliation{Department of Physics, University of Rome La Sapienza, P.le A. Moro 5, 00185 Roma, Italy}
\author{Florian Rasch}
\affiliation{Functional Nanomaterials, Institute for Materials Science, Kiel University, Kaiser Str. 2, 24143 Kiel, Germany}
\author{Martin R. Lohe}
\affiliation{Center for Advancing Electronics Dresden (CFAED) \& Department of Chemistry and Food Chemistry, Technische Universit\"{a}t Dresden, Helmholtzstra\ss e 10, 01069 Dresden, Germany}
\author{Ali Shaygan Nia}
\affiliation{Center for Advancing Electronics Dresden (CFAED) \& Department of Chemistry and Food Chemistry, Technische Universit\"{a}t Dresden, Helmholtzstra\ss e 10, 01069 Dresden, Germany}
\author{Fabian Sch\"{u}tt}
\affiliation{Functional Nanomaterials, Institute for Materials Science, Kiel University, Kaiser Str. 2, 24143 Kiel, Germany}
\author{Xinliang Feng}
\affiliation{Center for Advancing Electronics Dresden (CFAED) \& Department of Chemistry and Food Chemistry, Technische Universit\"{a}t Dresden, Helmholtzstra\ss e 10, 01069 Dresden, Germany}
\author{Rainer Adelung}
\affiliation{Functional Nanomaterials, Institute for Materials Science, Kiel University, Kaiser Str. 2, 24143 Kiel, Germany}
\author{Stefano Lupi}
\affiliation{Graphene Labs, Istituto Italiano di Tecnologia, Via Morego 30, 16163 Genova, Italy}
\affiliation{Department of Physics, University of Rome La Sapienza, P.le A. Moro 5, 00185 Roma, Italy}
\date{\today}
\maketitle
\indent \textbf{Graphene hydrophobic coatings paved the way towards a new generation of optoelectronic and fluidic devices. Nevertheless, such hydrophobic thin films rely only on graphene non-polar surface, rather than taking advantage of its surface roughness. Furthermore, graphene is typically not self-standing. Differently, carbon aerogels have high porosity, large effective surface area due to their surface roughness, and very low mass density, which make them a promising candidate as a super-hydrophobic material for novel technological applications. However, despite a few works reporting the general super-hydrophobic and lipophilic behavior of the carbon aerogels, a detailed characterization of their wetting properties is still missing, to date. Here, the wetting properties of graphene aerogels are demonstrated in detail. Without any chemical functionalization or patterning of their surface, the samples exhibit a super-lipophilic state and a stationary super-hydrophobic state with a contact angle up to $\mathbf{150\pm15^{\circ}}$ and low contact angle hysteresis $\mathbf{\approx15^{\circ}}$, owing to the fakir effect. In addition, the adhesion force of the graphene aerogels in contact with the water droplets and their surface tension are evaluated. For instance, the unique wettability and enhanced liquid absorption of the graphene aerogels can be exploited for reducing contamination from oil spills and chemical leakage accidents.}\\
\indent In general, the realization of artificial hydrophobic surfaces depends on the material surface chemical composition and its morphological structure. Although the chemical composition is an intrinsic material property, it can be engineered to decrease the solid surface tension \cite{Gennes2003}, therefore increase the hydrophobicity of the surface. On the other hand, surface roughness \cite{Feng2008,Wenzel1936} (micro- and nano-morphology) may also enhance the hydrophobicity, particularly by exploiting hierarchical \cite{Feng2008,XingJiuHuang2007,DeNicola2015b,DeNicola2015d,DeNicola2015a} and fractal architectures \cite{Shibuichi1996,DeNicola2015a}, that allow the formation of air pockets to prevent water imbibition. Nonetheless, the fabrication of a permanent super-hydrophobic surface is a challenging task. Recently, time durability \cite{DeNicola2015b,DeNicola2015d,DeNicola2015a}, chemical \cite{Wang2011}, mechanical \cite{Jung2009}, and thermal stability \cite{Wang2008} have been addressed. Among the numerous materials having the two aforementioned features, graphene offers versatility, stability, and multi-functionality owing to its unique optical \cite{Grigorenko2012} and electronic \cite{Banszerus2015} properties, making its usage widespread in hydrophobic surface realizations \cite{Wang2009,Rafiee2010,Shin2010}.
\section*{Introduction}
\label{sec:intro}
\indent Graphene is constituted by a $sp^{2}$ lattice of graphitic carbon, thus it has a slightly hydrophilic (graphite contact angle $\approx86^{\circ}$ \cite{Adamson1997}) but non-polar surface. Nevertheless, surface functionalization and substrate interactions may be exploited to tailor graphene wetting properties in a controlled fashion \cite{Rafiee2010,Shin2010}. However, graphene has been used so far to realize hydrophobic coatings only employing its non-polar surface, rather than taking advantage of its surface roughness. Furthermore, graphene is not self-standing in most application, acting thus as a coating.\\
\indent On the other hand, carbon aerogels \cite{Mecklenburg2012,Lin2011,Wu2014,Luo2017,Scarselli2015} are carbon-based, macroscopic, three-dimensional structures characterized by a randomly crosslinking network of hierarchical nanostructures and microstructures. Typically, aerogels have high porosity, very low mass density, and large effective surface area due to their hierarchical high surface roughness, which make them a promising candidate as a super-hydrophobic material for novel technological applications. Despite a few works \cite{Lin2011,Wu2014,Luo2017,Scarselli2015} reporting the observation of a super-hydrophobic and lipophilic behavior in carbon aerogels, the characterization of the wetting properties of such a class of materials is still missing, to date.\\
\indent Here, we give insight on the wetting properties of graphene aerogels in detail. In particular, we show that owing to their particular morphology and high surface porosity (up to 0.81), our graphene aerogels exhibit a super-lipophilic state and a stationary super-hydrophobic state, achieving high apparent contact angle values $\theta^{\star}\geq150^{\circ}$, with low contact angle hysteresis \cite{Gennes2003} (CAH) $\approx15^{\circ}$, and high work of adhesion (10 mJ/m$^{2}$). In addition, we characterized the graphene aerogel solid surface tension.
\section*{Results and discussion}
\label{sec:results}
\begin{figure}[t]
\centering
\includegraphics[width=0.5\textwidth]{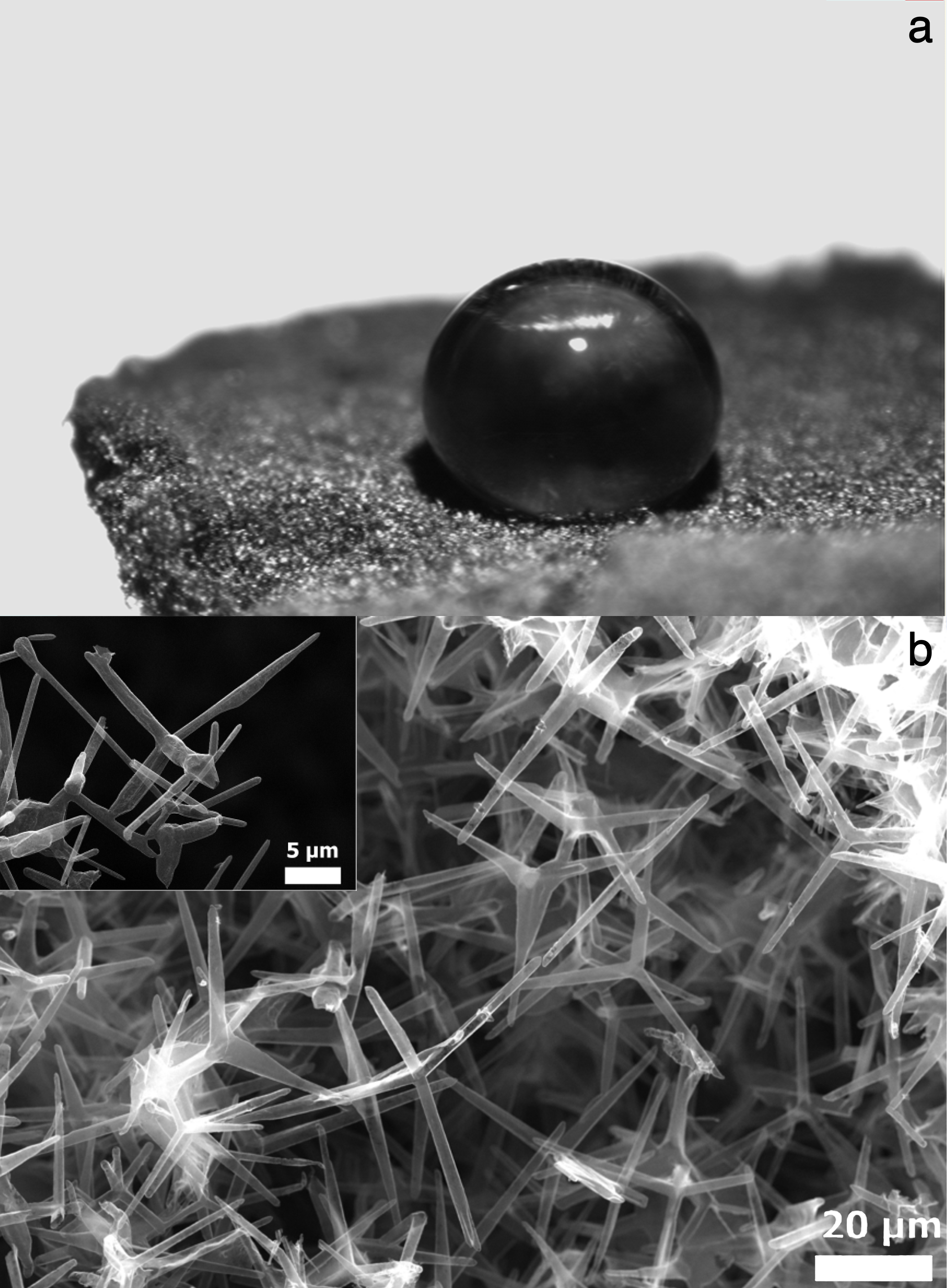}
\caption{Aerographene aerogels morphology. \textbf{a}, Image of a water droplet cast on a hydrophobic Aerographene aerogel. \textbf{b}, Representative SEM micrograph of the Aerographene aerogel surface. \textbf{Inset}, Detail of carbon tetrapods.}
\label{fig:Figure1}
\end{figure}

\indent The graphene aerogels studied are made of a cellular material called Aerographene (see Methods). The samples have a square-shaped surface with a lateral size $L=2$ cm, a thickness $d=0.2$ cm, and a mass $m=5-15$ mg. A representative sample is reported in Figure \ref{fig:Figure1}a. From the scanning electron microscopy (SEM) micrograph in Figure \ref{fig:Figure1}b, a random network made of micrometer-sized carbon tetrapods \cite{Mishra2013,Meija2017,Taale2019} constituting the aerogel surface (Figure \ref{fig:Figure1}b, inset) can be observed. This particular shape is due to the precursor ZnO template employed during the material synthesis. Therefore, such a highly porous (volume porosity up to $\approx0.99$) system cannot be considered as an homogeneous solid.\\
\begin{figure*}[ht]
\centering
\includegraphics[width=1\textwidth]{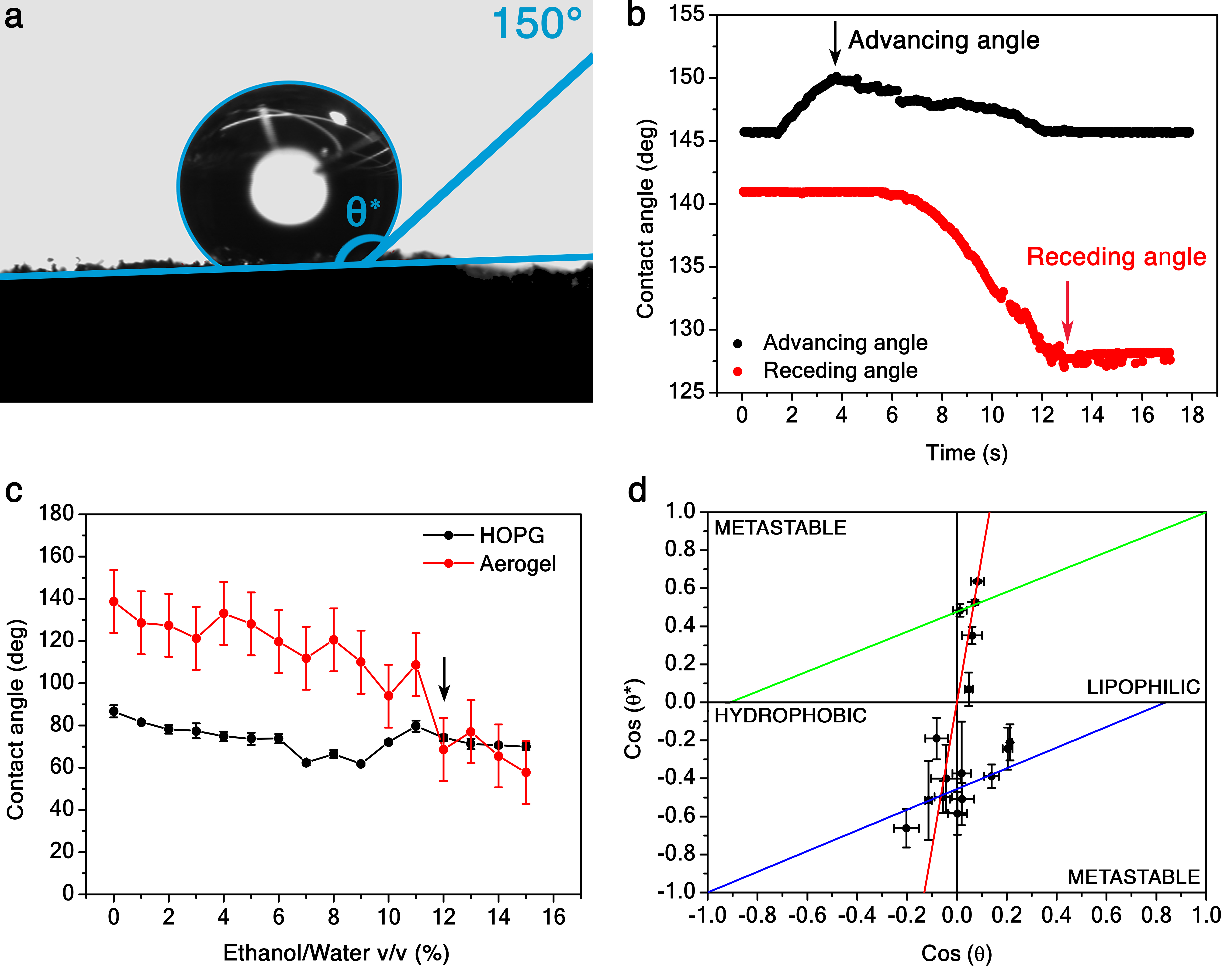}
\caption{Contact angle measurements. \textbf{a}, Image of a sessile water droplet cast on a super-hydrophobic ($\theta^{\ast}=150\pm15^{\circ}$) Aerographene aerogel. \textbf{b}, Contact angle as a function of time by increasing (black dots) and decreasing (red dots) the water droplet volume cast on the aerogel. The advancing and receding contact angles are marked. \textbf{c}, Contact angles of Aerographene aerogel (red dots) and HOPG (black dots) as a function of the ethanol volume concentration in water. The black arrow marks the lipophilic transition point between the Wenzel and Cassie-Baxter regime. \textbf{d}, Wenzel-Cassie-Baxter wetting phase diagram of the Aerographene aerogel surface with respect to the HOPG surface. Wetting states are studied changing the liquid surface tension by adding different concentrations in volume of ethanol in water. The fit of Wenzel equation (red solid line) reports a roughness factor $r=7.6\pm0.4$, while the fits of lipophilic (green solid line) and hydrophobic (blue solid line) Cassie-Baxter equations report respectively a liquid fraction $\phi_{+}=0.48\pm0.03$ and an air fraction $\phi_{-}=0.45\pm0.02$. The Wenzel-Cassie-Baxter transition point in the hydrophobic regime is the intersection between the red and blue solid lines, while in the lipophilic regime it is the intersection between the red and green solid lines. Error bars are standard deviations of data measured over different sample areas.}
\label{fig:Figure2}
\end{figure*}
\indent The mass density of the Aerographene aerogels can be estimated at first approximation by $\rho=m/V$. However, this is the total mass density, as aerogels are matter with a mixed phase of gas and solid. Hence, a better estimation of the solid mass density can be provided by the effective medium approximation $\rho=\rho_{c}(1-\Phi_{-})+\rho_{air}\Phi_{-}$, with $\Phi_{c}+\Phi_{-}=1$, where $\rho_{c}=2200$ kg/m$^{3}$ and $\Phi_{c}=V_{c}/V$ are respectively the density and the volume fraction of the carbon phase, while $\rho_{air}=1.225$ kg/m$^{3}$ and $\Phi_{-}=V_{air}/V$ are the density and the volume fraction of air, respectively. Hence, the effective mass density is defined as $\rho_{eff}=\rho_{c}(1-\Phi_{-})$. Therefore, the aerogel porous fraction is $\Phi_{-}=0.992-0.998$, leading to an effective mass density down to $\rho_{eff}=5$ kg/m$^{3}$.\\
\indent Owing to their random surface morphology, the static contact angle on porous media can be defined only on average \cite{Gennes2003}. The contact angle of a water droplet cast on a representative Aerographene aerogel is shown in Figure \ref{fig:Figure2}a. We measured a maximum apparent contact angle $\theta^{\ast}=150\pm15^{\circ}$ on our samples, thus the sample is super-hydrophobic ($\theta^{\ast}\geq150^{\circ}$). The error on the contact angle was evaluated by the CAH, that is defined as the difference between the advancing and receding contact angle (Figure \ref{fig:Figure2}b). Such a low CAH is due to the highly rough and porous surface of the samples and it is peculiar of the fakir effect \cite{Quere2002}.\\
\indent Moreover, we evaluated the work of adhesion of the aerogel surface in contact with water by the Young-Dupr\'e equation \cite{Gennes2003}
\begin{equation}
W_{adh}=\gamma_{LV}(1+\cos{\theta^{\ast}}).
\end{equation}
For the super-hydrophobic sample in Figure \ref{fig:Figure2}a, $W_{adh}\approx10$ mJ/m$^{2}$. Therefore, for a water drop with diameter 1 mm, the adhesion force of the Aerographene aerogel in contact with the drop is $F_{adh}\approx10$ $\mu$N. For instance, the obtained result is about 17\% lower than the adhesion force of a single gecko foot-hair \cite{Autumn2000}, but 10 times higher than that of \textit{Salvinia} leaf \cite{Hunt2011}, and compatible with that reported for carbon nanotube films \cite{DeNicola2015d}.\\
\indent In order to better understand the wettability of Aerographene aerogels, we characterized their wetting states with respect to a highly oriented pyrolytic graphite (HOPG) reference substrate (Bruker). In Figure \ref{fig:Figure2}c, we report the contact angle of the super-hydrophobic aerogel and the HOPG as a function of the concentration in volume of ethanol in water. Since ethanol has a lower liquid-vapor surface tension ($\gamma_{LV}=22$ mN/m) than water ($\gamma_{LV}=72$ mN/m), the higher the ethanol concentration, the lower the surface tension of the overall solution.\\
\indent Generally, the contact angle between a chemically homogeneous solid surface and a liquid droplet obeys to the Young relation \cite{Gennes2003}
\begin{equation}
\cos{\theta}=\frac{\gamma_{SV}-\gamma_{SL}}{\gamma_{LV}},
\end{equation}
where $\gamma_{SV}$ and $\gamma_{SL}$ are the solid-vapor and solid-liquid surface tensions, respectively. Therefore, the lower the surface tension of the liquid droplet, the lower the Young contact angle of the HOPG. This phenomenon is connected to the lipophilicity of the non-polar surface of carbon-based materials. Indeed, a minimum contact angle $(\theta=70^{\circ})$ can be measured on the HOPG for pure ethanol droplets (Figure \ref{fig:Figure2}c). On the other hand, it was not possible to measure the aerogel contact angle with acetone, ethanol, and glycerol because the material instantly and completely absorbed the droplets. Hence, the Aerographene aerogels are super-lipophilic $(\theta^{\ast}<5^{\circ})$ as a consequence of their super-hydrophobic behavior. We further observed that for $\theta=77^{\circ}$, $(\cos{\theta^{\ast}}=0.2)$ there is an intersection point between the two curves in Figure \ref{fig:Figure2}c, beyond which the aerogel surface becomes more lipophilic than the graphite surface. That point corresponds to the transition point from the Wenzel to the Cassie-Baxter state in the lipophilic regime of the Wenzel-Cassie-Baxter phase diagram \cite{DeNicola2015b}, as confirmed from the plot (first quadrant) in Figure \ref{fig:Figure2}d. We fit our data with the lipophilic Cassie-Baxter equation \cite{Cassie1944,DeNicola2015a}
\begin{equation}
\cos{\theta^{\ast}}=\left(1-\phi_{+}\right)\cos{\theta}+\phi_{+}, \qquad 1=\phi+\phi_{+},
\label{eq:cassiehydrophilic}
\end{equation} 
by considering the Aerographene aerogel as a composite surface with a solid surface fraction $\phi$, a surface fraction wetted by the liquid $\phi_{+}$, an aerogel contact angle $\theta^{\ast}$, and a HOPG Young contact angle $\theta$. We obtained from the fit (green solid line) a liquid fraction $\phi_{+}=0.48\pm0.03$ in contact with the droplet. We point out that metastable \cite{Giacomello2012} Cassie-Baxter states coexist with Wenzel states, which are stable as they are lower in surface free energy \cite{DeNicola2015b}. Therefore, we fit our data in Figure \ref{fig:Figure2}d with the Wenzel equation \cite{Wenzel1936}
\begin{equation}
\cos{\theta^{\ast}=r\cos{\theta}}, \qquad r\geq1,
\label{eq:wenzel}
\end{equation}
where $r$ is the roughness factor (i.e, the ratio between the actual wet surface area and its geometrical projection on the plane) \cite{DeNicola2015b}. The fit (red solid line) returned $r=7.6\pm0.4$, meaning that the Aerographene aerogel surface is quite rough (the root-mean-squared roughness is $\approx3$ $\mu$m). In addition, from Equation \ref{eq:cassiehydrophilic} and \ref{eq:wenzel} we obtain the relation
\begin{equation}
\cos{\theta}=\frac{\phi_{+}}{r+\phi_{+}-1},
\end{equation}
from which we infer that the lipophilic Wenzel/Cassie-Baxter transition point occurs at $\cos{\theta}=0.07$ (the intersection between the red and green solid lines in Figure \ref{fig:Figure2}d), hence confirming that the achieved lipophilic Cassie-Baxter states are metastable.\\
\indent On the other hand, in the hydrophobic regime (third quadrant of the plot) we observe a continuous transition between the Wenzel and the Cassie-Baxter states beyond $\cos{\theta}=0$. Also, the plot depicts that the transition occurs by passing through metastable states extending into the fourth quadrant, slowing down the dewetting process. By fitting our data (blue solid line) in Figure \ref{fig:Figure2}d with the hydrophobic Cassie-Baxter equation \cite{Cassie1944,DeNicola2015a}
\begin{equation}
\cos{\theta^{\ast}}=\left(1-\phi_{-}\right)\cos{\theta}-\phi_{-}, \qquad 1=\phi+\phi_{-},
\label{eq:cassiehydrophobic}
\end{equation} 
we obtained an air surface fraction $\phi_{-}=0.45\pm0.02$ below the liquid droplet. Furthermore, from Equation \ref{eq:cassiehydrophobic} and \ref{eq:wenzel} we obtain the relation
\begin{equation}
\cos{\theta}=\frac{\phi_{-}}{1-r-\phi_{-}},
\end{equation}
from which we infer that the hydrophobic Wenzel/Cassie-Baxter transition point is at $\cos{\theta}=-0.07$ (the intersection between the red and blue solid lines in Figure \ref{fig:Figure2}d), thus the maximum hydrophobic Cassie-Baxter state achieved is not metastable. These results suggest an air pocket formation. Therefore, we assert that the reason of the improved hydrophobicity/lipophilicity of the Aerographene aerogel over HOPG, is the fakir effect (high contact angle and low CAH) induced by the aerogel microstructure. When the interaction of the surface with the liquid is hydrophobic, the particular surface morphology promotes the air pocket formation. Otherwise, when the interaction between the surface and the liquid is lipophilic, the morphology induces the development of a precursor liquid film \cite{Gennes2003}, improving the wetting behavior of the Aerographene aerogel surface. Furthermore, in composite rough surfaces, their morphology may favor a wetting transition from a Wenzel to a Cassie-Baxter state, due to air trapping \cite{DeNicola2015b}. This transition typically occurs by thermodynamically metastable states \cite{Giacomello2012}.\\
\begin{figure}[t]
\centering
\includegraphics[width=0.5\textwidth]{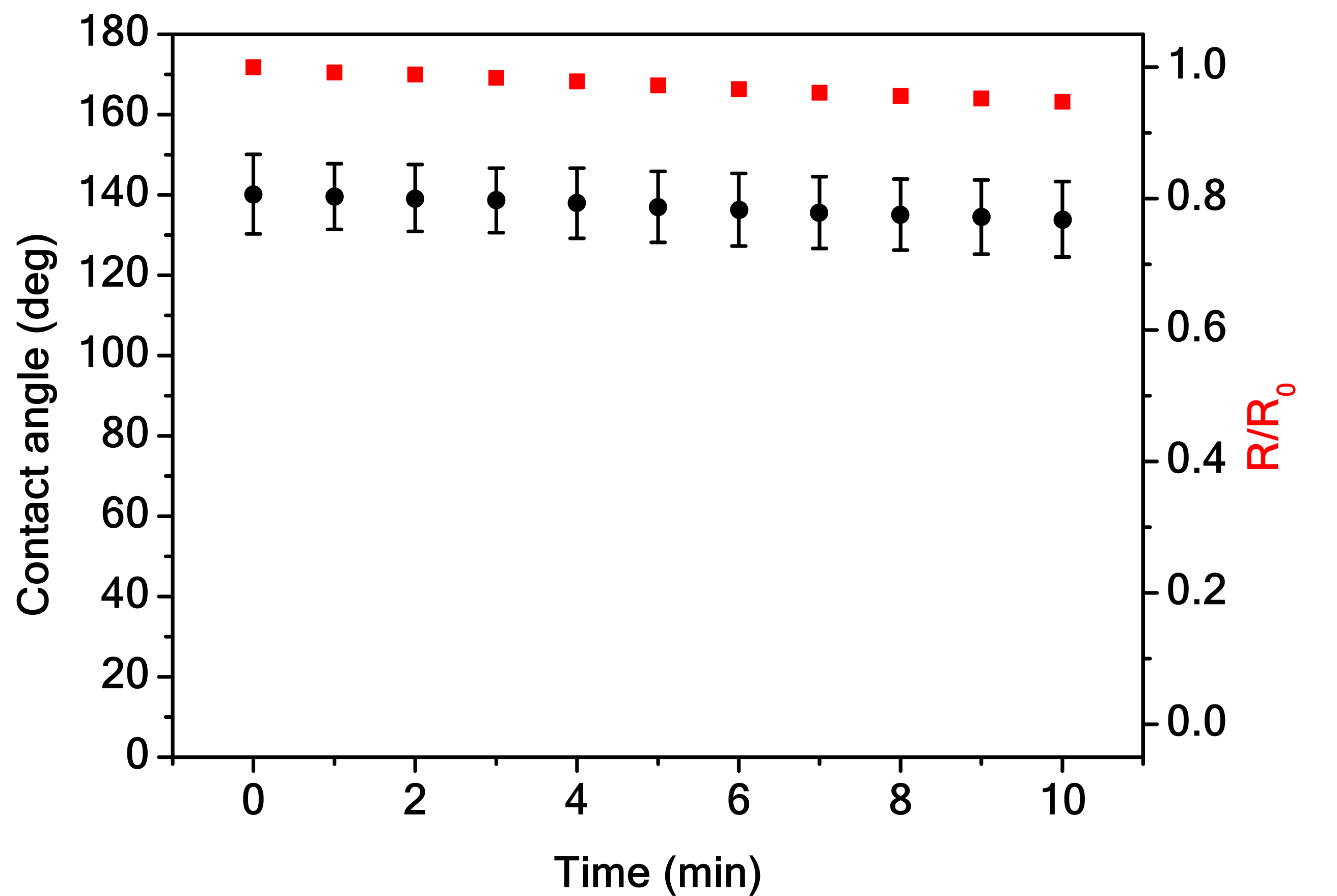}
\caption{Wetting stability. Stability of the contact angle in the super-hydrophobic regime over time (black dots). Normalized radius of the water droplet as a function of the elapsed time (red squares). $R_{0}$ is the radius initial value.}
\label{fig:Figure3}
\end{figure}
\indent In addition, we investigated the stability of the Aerographene aerogel hydrophobicity over time. Figure \ref{fig:Figure3} reports the variations of the contact angle value as a function of the elapsed time since water was cast on the aerogel. In such an experiment, we show that despite the samples are porous, the contact angle is constant up to 10 min. The slight linear decrease of the droplet radius in time is only due to the liquid evaporation and not to the suction by the aerogel, otherwise the contact angle would not be constant in time along with droplet radius \cite{Gennes2003,DeNicola2015a}. In addition, we demonstrated a remarkable stability over several hours of the hydrophobic Cassie-Baxter state for the Aerographene aerogel (Supplementary Material). This result is particularly remarkable, as the water contact angle of other carbon-based surfaces has been reported to decrease linearly with time, from an initial value of $146^{\circ}$ to 0 within 15 min \cite{Huang2005}.\\
\indent Moreover, we estimated the solid surface tension $\gamma_{SV}$ of the HOPG substrate (Figure \ref{fig:Figure4}a) and the graphene aerogel (Figure \ref{fig:Figure4}b) by the Neumann equation \cite{Adamson1997}
\begin{equation}
\log\left[\gamma_{LV}\left(\frac{1+\cos{\theta^{\ast}}}{2}\right)^{2}\right]=\log\left(\gamma_{SV}\right)-\beta\left(\gamma_{LV}-\gamma_{SV}\right)^{2},
\end{equation}
where $\gamma_{LV}$ is the liquid-vapor surface tension of the different concentrations in volume of ethanol
in water and $\beta$ is a constant. The fit in the lipophilic regime returns $\gamma_{SV}\approx42$ mN/m and $\gamma_{SV}\approx63$ mN/m for the HOPG and the graphene aerogel, respectively.\\
\indent In summary, we investigated the wetting properties of graphene aerogels. We demonstrated that such a material can be in a super-lipophilic, thus super-hydrophobic state that is stationary. The evaluated adhesion force of the water droplets with the aerogel is higher than the leaves of aquatic plants and we estimated the graphene aerogel solid surface tension. Stationary super-hydrophobic behavior of solid surfaces is a relevant property in a number of natural \cite{Feng2008} and technological processes \cite{Furstner2005} with several industrial applications such as waterproof surfaces \cite{Lau2003}, anti-sticking \cite{Wang2007}, anti-contamination \cite{XingJiuHuang2007}, self-cleaning \cite{Furstner2005}, anti-fouling \cite{Zhang2005}, anti-fogging \cite{Lai2012}, low-friction coatings \cite{Jung2009}, adsorption \cite{Li2010a}, lubrication \cite{Adamson1997}, dispersion \cite{Gennes2003}, self-assembly \cite{DeNicola2015b}, and optoelectronic and fluidic devices \cite{Fang2015,Min2011}.
\begin{figure}[t]
\centering
\includegraphics[width=0.5\textwidth]{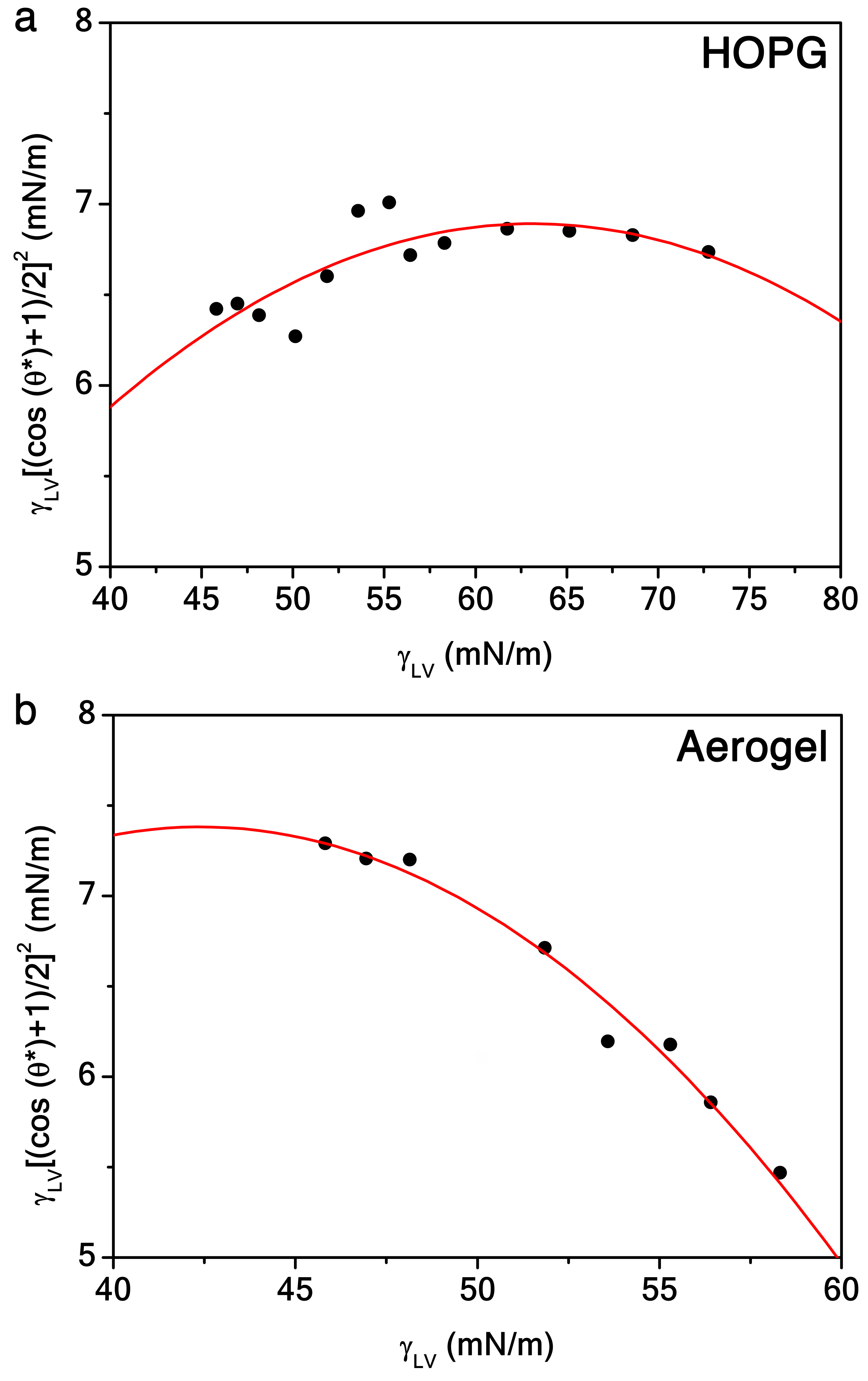}
\caption{Solid surface tension characterization. Neumann equation in the lipophilic regime as a function of the liquid-vapor surface tension of different concentrations in volume of ethanol in water for the HOPG substrate (\textbf{a}) and the graphene aerogel (\textbf{b}). Red solid curves are quadratic polynomial fits.}
\label{fig:Figure4}
\end{figure}
\section*{Methods}
\indent \textbf{Fabrication of the Aerographene aerogels.} A macroscopic ($2\times2\times0.5$ cm) and highly porous ($\approx0.94$) template consisting of interconnected ceramic ZnO tetrapods \cite{Mishra2013,Meija2017,Taale2019} was infiltrated with an aqueous dispersion (2 mg/ml) of electrochemically exfoliated graphene flakes \cite{Parvez2014}. After evaporation of the solvent and self-assembly of the graphene flakes on the template surface, the ZnO template was removed by chemical etching in 1M HCl and subsequent critical point drying resulted in a macroscopic, free-standing random network of hollow, interconnected carbon tetrapods having a wall thickness on the order of a few nanometers, length $\approx20$ $\mu$m, and diameter about a few micrometers.\\
\indent \textbf{Contact angle measurements.} Images of sessile water drops deposited on Aerographene aerogels were acquired by Dataphysics OCA instrument and analyzed by its software. In order to estimate the average, maximum, and minimum contact angle on the aerogel surface, static, advancing, and receding contact angles \cite{Gennes2003} were measured, respectively, by increasing and decreasing the volume of the drop by 1 $\mu$L steps. In static contact angle measurements, the volume of the deionized (18.2 M$\Omega$cm) water droplet was $V=15$ $\mu$L. Every contact angle was measured 15 s after drop casting to ensure that the droplet had reached its equilibrium position, and it was averaged over the values obtained in different areas of the sample surface. Due to the heterogeneous and porous aerogel surface, we evaluated the error on the contact angle values by $CAH\approx15^{\circ}$. The experimental contact angle probably corresponds to a lower limit, as the water droplet might cause some small dimple on the aerogel surface. The calculated surface air and liquid fractions are effective values, as they were determined by macroscopic techniques relying on the surface properties of the material. The liquid-vapor surface tension of the concentrations in volume of ethanol in water was measured by the pendant drop method \cite{Gennes2003}.

\begin{acknowledgments}
The authors acknowledge that this project has received funding from the European Union’s Horizon 2020 research and innovation programme under grant agreement No. 785219 - GrapheneCore2. The authors also acknowledge that this project has received the financial support of the Bilateral Cooperation Agreement between Italy and China of the Italian Ministry of Foreign Affairs and of the International Cooperation (MAECI) and the National Natural Science Foundation of China (NSFC), in the framework of the project of major relevance 3-Dimensional Graphene: Applications in Catalysis, Photoacoustics and Plasmonics. Also, this work was supported by the Progetto FISR - C.N.R. "Tecnopolo di nanotecnologia e fotonica per la medicina di precisione" - CUP B83B17000010001.
\end{acknowledgments}
\section*{Author contribution}
\label{sec:AuthorContribution}
\noindent F.D.N., I.V., L.D.T., F.R., F.S., R.A., M.R.L., A.S.N., X.F., and S.L. conceived the experiments. M.R.L. and A.S.N. prepared the graphene dispersion and F.R. and F.S. fabricated the samples and acquired the SEM images. F.D.N., I.V, and L.D.T. performed the wetting characterization of the samples and data analysis. All the authors discussed the experimental implementation, the results, and contributed in writing the paper.\\\\
The authors declare no competing interests.
\end{document}